# A Static Distributed-parameter Circuit Model to Study Electrical Stimulation on Muscle Tissue

Jiahui Wang[1,2,3], Hao Wang[1,2,3]*, Xin Yuan Thow[2], Nitish V. Thakor[1,2], Shih-Cheng Yen[1,2], Chengkuo Lee[1,2,3]*

[1] Department of Electrical & Computer Engineering, National University of Singapore, 4 Engineering Drive 3, 117576, Singapore

[2] Singapore Institute for Neurotechnology (SINAPSE), National University of Singapore, 28 Medical Drive, #05-COR, 117456, Singapore

[3] Center for Intelligent Sensors and MEMS, National University of Singapore, 4 Engineering Drive 3, 117576, Singapore

**Abstract**

The existing Finite Element Modeling (FEM) fails to model the real electric field that activates a neuron, because the computation capability is insufficient to represent the neural tissue down to an atom-level. Thus, to reveal the real electrode-tissue interactions, we adopt the idea of using a circuit to simulate voltage waveforms, as proposed in the Circuit-Probability theory (C-P theory). Here we show a distributed-parameter circuit model to systematically study how the interaction between electrode and tissue is affected by the electrode position, input current waveform, and the biological structures in the muscle. Our model explains and predicts various phenomena in muscle stimulation, guides new stimulation electrode and method design, and more importantly, facilitates a fundamental understanding of the physical process during electrode-tissue interaction. With some proper modifications, other neural tissues, including the Peripheral Nervous System (PNS) and the Central Nervous System (CNS), can also be studied with this distributed-parameter circuit model.

Electroceuticals, where electrical stimulation is delivered to the nervous system and muscles, are becoming widespread therapeutic solutions to people with neurological disabilities. For example, people with spinal cord injury (SCI) above the sixth cervical vertebra are unable to control extant limbs due to interruption to the motor pathway[1]. Functional electrical stimulation (FES) could benefit these people by restoring functional actions, like voluntary grasp, via the electrical stimulation of specific muscles. However, despite its application and medical promise, the electrophysiology of electrical stimulation is neither precise nor well understood. The history of electrophysiology dates back to 1770, when Luigi Galvani first discovered bioelectricity as he made a frog muscle twitch by accidentally creating a battery from surgical instruments[2]. The mechanism of electrical stimulation remained elusive until 1952, when Alan Hodgkin and Andrew Huxley proposed a quantitative description of membrane current on the unmyelinated squid giant axon[3]. In 1976, Donald McNeal first applied the established nerve axon models to explain excitation of nerve tissue, by bringing in the concept of shared electric field between the stimulation electrodes and the excitable nerve tissue[4]. With increased computing power, modern computational models further improved McNeal's method, by accounting for additional parameters, such as the anisotropic extracellular conductivity in electric field simulation and non-linear response of neuronal cells and axons. Such modern computational models include the field-neuron model, which has been applied to analyze electrical stimulation on the peripheral nervous system[5][6][7] and the central nervous system[8][9].

Despite the development of these computational models, there are large discrepancies between these computational models and experimental observations. Firstly, the tissue should not be considered as purely resistive[5-9], as recent indirect[10][11][12] and direct[13] evidence suggests that the extracellular medium consists of non-resistive components. Second, the deterministic gating property assumed in the computational models[5-9] contradicts the probabilistic gating property as observed in single ion channel recording[14-18]. In addition, there lacks a complete model that can systematically study how the interaction between the electrode and tissue is affected by the electrode position, input current waveforms, and the biological structures of the neural tissue. In this paper, we aim to achieve such a complete model that explains and predicts all various phenomena in muscle stimulation.

In the circuit-probability theory (C-P theory) proposed in our previous submission[19], the composite biological structures in the tissue are neglected for circuit simplification. In this study, by considering motoneurons, muscle fibers and extracellular medium, the lumped-parameter circuit adopted in the C-P theory is now expanded to a distributed-parameter circuit, enabling a more detailed investigation of the interaction between stimulating electrode sites and excitable tissues. This muscle model can qualitatively explain many phenomena:

1. With different locations of the stimulation electrode sites relative to the target motoneuron (motoneuron-electrode position), the stimulation efficiency can either increase or decrease with respect to the spacing between the two electrode sites.

2. The motoneuron-electrode position determines the polarity of the cross-membrane voltage waveform, which is an unsolved issue using the lumped-parameter circuit in the C-P theory.

3. The force mapping curves measured with increasing current amplitude may form a certain shape with multiple curvatures, which has been observed in our previous experiments as well. A conventional explanation for this phenomenon is that multiple groups of motoneurons are sequentially recruited when the current amplitude is increased[20][21]. But in our model, this phenomenon can be theoretically derived and numerically calculated.

4. Instead of forming a force mapping curve (a measured curve of the generated force with respect to input current amplitude or pulse width) with multiple curvatures, we observe that the recruitment of multiple motoneuron groups will induce an unstable output force at the transition current, showing an abnormally high error bar in the force mapping curve. This phenomenon is reflected in the measurement data. It helps us to tell when an additional group of motoneurons is recruited during the experiment.

Based on a comprehensive understanding of muscle stimulation, we propose an effective design of the stimulation electrode and a corresponding stimulation method to reduce muscle fatigue and control stimulation efficiency. Firstly, a double-side multiple-channel polyimide electrode design is proposed to reduce stimulation fatigue by alternatively activating the motoneurons on the frontside and backside of the electrode. Secondly, a comprehensive calibration of the force mapping curve using different combinations of the electrode sites is necessary because the stimulation efficiency and linearity of force control using a specific electrode pair is determined by the motoneuron-electrode location, meaning that a universal parametric force control method does not exist.

In addition, we propose new evidence to support the idea of inductive myelin sheath on motoneuron axons, which is claimed in the C-P theory. In this study, we provide a theoretical and experimental exploration of the excitability differences between the myelinated motoneurons and unmyelinated muscle fibers.

**Circuit-Probability simulation concept with distributed-parameter circuit**

In the previous C-P theory, the target motoneuron is modeled as a parallel RLC circuit. The extracellular medium is simplified as parallel impedance and integrated within the leakage resistor of the parallel RLC circuit. In this paper, the entire muscle tissue, including the target motoneurons and extracellular tissue, is modeled as a distributed-parameter circuit network (Fig. 1). The detailed circuit configuration of each block in this network is determined by its biological features. The myelinated motoneurons are modeled as parallel RLC components (green block). The unmyelinated muscle fibers are modeled as RC components (blue block). The extracellular medium is represented with pure resistive components ($R_s$) connecting each block.

This circuit network can be further expanded by adding more functional components (e.g. fat tissue and skin as the peripheral circuits) and revised when studying different experimental scenarios (e.g. different configurations and implantation positions of the stimulating electrodes). In this study, a home-made double-side multiple-channel polyimide electrode was implanted in the muscle belly, transversal to the muscle fibers of the Tibialis Anterior (TA) muscle in all experiments, and the force generated by stimulation was measured by a force gauge tied to the ankle (Fig. 2a). Thus, the corresponding circuit network is revised as Fig. 2b. Each electrode site (e1 to e6 refer to the frontside; e1' and e6' refer to the backside) is connected to a node (E1 and E6 refer to the frontside; E1' and E6' refer to the backside). The non-conducting polyimide layer is modeled as broken connections between the frontside and backside. By connecting a current source to an arbitrary electrode sites and a voltage meter to the capacitor in an arbitrary block, the effective voltage waveform at any position in the tissue can be calculated. This voltage waveform will then be used for a probability calculation with a set of parameters, $\alpha$, $\beta$, and $V_{Threshold}$ (parameters to calculate the probability of initiating an action potential), which is the same as the C-P theory.

**Results**

**Influence of motoneuron-electrode position on voltage waveforms.** Apparently, the exact voltage waveform on each block in Fig. 2b is determined by the block position and the electrode sites selected to deliver current. This motoneuron-electrode position will affect the amplitude, shape and the polarity of the voltage waveform on each block. Since there are many motoneuron-electrode combinations, two simplified situations are simulated (Fig. 3a and Fig. 3b), to qualitatively investigate the effect of this motoneuron-electrode position upon the voltage waveform.

In Fig. 3a, fixed electrode sites (positive electrode site between P3 and P4; negative electrode site between P9 and P10) are selected to deliver negative-first biphasic square current input. Voltage waveforms upon different blocks (P1 to P12) are modeled and shown in Fig. 3c. For these different blocks, the voltage amplitude changes, and the polarity also switches. The blocks which are next to the electrode sites (P3 and P4; P9 and P10) have the largest voltage amplitude. Meanwhile, the voltage polarity gradually changes from negative-first to positive-first when the position changes from P1 to P12.

In another situation (Fig. 3b), two blocks are fixed (P1 and P2) while the spacing between the electrode sites is increased (positive electrode site between P1 and P2 is fixed; negative electrode site changes from E1 to E8). The voltage waveforms on P1 and P2 (Fig. 3d and Fig. 3d') show an opposite changing trend. The peak amplitude of the block outside the electrode sites (P1) and between the electrode sites (P2) increases and decreases with the increasing electrode spacing, respectively.

The changing trends of the peak amplitude and polarity (green lines in Fig. 3c, Fig. 3d and Fig. 3d') agree with the results using conventional method of electric field distribution simulation. However, our model can also reveal information in the time domain: voltage waveform, which is critical in determining neuron activation under the C-P theory, but unavailable in previous methods. This voltage waveform can help us understand more phenomena, which will be discussed in the next session.

**Influence of voltage waveform on force mapping curve: An explanation of the multi-curvature phenomenon.** For muscle stimulation using square wave current, it is widely observed that the force mapping curves measured with increasing current amplitude may show multiple curvatures[20][21]. A conventional explanation for this phenomenon is that multiple groups of motoneurons are sequentially recruited when current amplitude is increased. Alternatively, this phenomenon can be quantitatively derived from the C-P theory. Using the same circuit configuration as Fig. 3a, the voltage waveforms of block P1 to P12 in Fig. 3c are shown in Fig. 4a and Fig. 4b, along with an estimated $V_{Threshold}$. Due to the voltage oscillation induced by RLC components, multiple effective voltage areas (indicated as A and B in Fig. 4a) will sequentially exceed $V_{Threshold}$ when current amplitude is increased, resulting in multiple curvatures in the probability mapping curves (Fig. 4a'). But this phenomenon will not always happen. For the voltage waveforms with only one area exceeding voltage threshold (indicated as C in Fig. 4b), their corresponding probability mapping curves will show a single curvature (Fig. 4b'). These two types of force mapping curves, with and without multiple curvatures, were observed in experiments.

Two conclusions can be summarized from the above modeling results and validated by force mapping results in two in vivo experiments (Fig. 4c, Fig. 4c'). First, due to the unknown motoneuron-electrode position after electrode implantation, the stimulating efficiency using the same electrode pair will vary in different experimental trials. In Fig. 4c, the electrode sites with the largest spacing (e1and e6) show minimum stimulation efficiency. However, in another experiment (Fig. 4c'), electrode sites of e1 and e6 show medium stimulation efficiency. Second, the multi-curvature phenomenon will not always happen in the force mapping curves. Whether the curve will have multiple curvatures is determined by the shape of the voltage waveform. The blue curve in Fig. 4c and purple curve in Fig. 4c' clearly show multi-curvature pattern, while other curves don't show such pattern.

**Activation of multiple groups of motoneurons.** As explained above, in contrast to the conventional explanation of multi-curvature force mapping curve as a sequential activation of multiple groups of motoneurons, the activation of a single group of motoneurons can already account for this phenomenon in our theory. We find that the unstable force output during transition periods of additional motorneuron group activation better characterize the recruitment of motornueron groups due to attenuation of voltage over a distance.

Two measured force mapping curves (e2e5, orange and e3e4, purple) are selected from Fig. 4c' and plotted in Fig. 5c (blue and orange). There is a sudden force increment accompanied with an abnormally large error bar for the blue curve. We believe that this error bar is a sign of an additional group of motoneurons starting to be recruited, due to the differing voltage acting on spaced out groups of motorneurons. In our experiments (Fig. 1a), the force measured is the summed-up force generated by multiple groups of activated motoneurons. When an additional group of motoneurons starts to be recruited in the stimulation, while the total force will be higher than activating one group alone, the force generated by this second group of motoneurons is still low at this stage. Furthermore, motorneurons that require higher stimulation current to be recruited are necessarily further away and the stimulation strength (voltage upon the motoneurons) will be low, resulting in unstable activation and force generated. Hence, the measured force curve will show a large error bar at this current. Then, when the current further increases, the stimulation strength of this second group of motoneurons will be higher to generate a stable force output, making the error bar recover to a normal level. This explains why this abnormal high error bar only occurs within a certain current range. Fig. 5d shows the force profile measured at this transition range (1000 µA to 1200 µA) of the blue curve in Fig. 5c. When current is high enough (1200 µA), the force profile will recover to a stable condition.

In addition to the measured force mapping curves, a good curve fitting with distributed-parameter circuit modeling can help us to understand the spatial distribution of these motoneurons within the muscle. Here a simple modeling case is demonstrated. The corresponding electrode pairs of these two curves are modeled in the circuit network (e3e4 in Fig.2b as E1+E1- in Fig. 5a; e2e5 in Fig. 2b as E2+E2- in Fig. 5a). The locations of the two target motoneurons (P1 and P2) are captured to fit the force mapping curves. The probability curves (the simulated curves of activation probability under the C-P theory with respect to input current amplitude or pulse width) are shown in Fig. 5b and Fig. 5b'. At low stimulation current, both E1 and E2 electrode pairs can only stimulate P1. However, when current increases to around 1600 µA, E2 starts to stimulate another motoneuron P2, while E1 has no influence on P2. From this modeling, the relative locations of the two motoneurons with respect to the electrode sites can be roughly estimated. This simple case demonstration shows the potential to apply this model for an in-depth investigation of the biological structure in the future.

**Influence of input current waveform polarity on motoneuron activation.** As shown in Fig. 3c, the polarity of the voltage waveforms at different positions will not be the same. This polarity is not only determined by its position and the polarity of the applied current, but also determined by which electrode is connected to the positive terminal of the current source. It means if the polarity of the input current and the connection of the electrode sites to current source are both reversed, the voltage waveform measured on the specific block will remain the same. One case by modeling is shown in Fig.6. Two electrode sites (E1 and E2) are connected to the current source, which can deliver monophasic square wave current pulses of different pulse widths. The voltage waveform by applying positive pulses from E1 and negative pulses from E2 are the same (Fig. 6b). Similarly, the voltage waveform by applying negative pulses from E1 and positive pulses from E2 are the same. Although the voltage waveforms in Fig. 6b and Fig. 6b' are just of the opposite polarity, the changing trend of the effective voltage area by increasing the pulse width is completely different.

Force mapping curves measured in two in-vivo experiments are consistent with the simulation results (Fig. 6c, Fig. 6c'). One force mapping curve was measured at small pulse width range (100-300 µs), and the other is measured at large pulse width range (100-800 µs). Two electrode sites (e1 and e6) of the furthest distance on our polyimide electrode were used. The force mapping curves are the same when e1 delivers positive current or e6 delivers negative current (blue and orange curves, with Fig. 6b' type activation). The recruitment curves are also the same when e1 delivers negative current or e6 delivers positive current (purple and green curves, with Fig. 6b type activation). In each measurement, the four recruitment curves form two groups of different changing trend with increasing pulse width. The flipping of voltage waveform polarity (Fig. 6b and Fig. 6b') causes the voltage waveforms to exceed the threshold in different patterns, so that the two groups have totally different changing trend with increasing pulse width.

Thus, for stimulations using two-electrode configurations, it is important to point out the polarity of current input together with which electrode site is connected to the positive terminal, as even when the same current input is used, results will be different when positive electrode is changed. This issue may not be so prominent in muscle stimulation, because the difference of using different positive electrode only shows in the stimulation efficiency. However, when it comes to the sciatic nerve stimulation using cuff electrode, different positive electrode may stimulate nerve fibers innervating different muscles with the same current input. This will lead to totally different results of activating different muscle groups. Thus, to get consistent results in the experiments, it is important to pay attention to both the polarity of current input and the chosen positive electrode.

**Independent activation of motoneurons using two-side polyimide electrode.** In Fig. 7a, due to the non-conducting polyimide layer, there will be broken connections between the frontside and backside of the electrode. Thus, when current is applied from the frontside electrode sites, motoneurons on the backside won't be activated. In other words, the electrode sites on one side can only cause extremely limited motoneuron activations on the other side of the electrode.

A modeling demonstration is shown in Fig. 7a-c. Two electrode sites (E+ and E-) on the frontside of the electrode are selected for current delivery. F1, F2, F3 are three motoneuron positions on the frontside, while B1, B2, B3 are three motoneuron positions on the backside of the electrode. The overall voltage amplitude on the frontside motoneurons is much larger than the backside motoneurons (Fig. 7b, Fig. 7b'). The probability curves show that only with much larger current, the backside motoneurons can be activated by the frontside electrode sites (Fig. 7c).

In-vivo experiments proved the independent activation of motoneurons using two-side polyimide electrode. At the beginning of the experiment, force mapping curves for both frontside and backside electrode sites were measured (blue curves in Fig. 7d and Fig. 7d'). Then, the frontside electrode sites were used for a long-time electrical stimulation to induce fatigue. The stimulation lasted for 2.5 min, and the force profile can be found in Fig. 7e. This force profile showed signs of muscle fatigue at the end of the 2.5 min stimulation, as the force dropped to half of the initial value. Right after this 2.5 min stimulation, the force mapping curve using frontside electrode sites was measured (orange curve in Fig. 7d). At that moment, because of the muscle fatigue on the frontside, the force mapping curve was much lower than the one measured in the beginning. Then, the force mapping curve using backside electrode sites was also measured (orange curve in Fig. 7d'). It was only slightly lower than the one measured in the beginning. Lastly, after 5 min resting, the force mapping curve using the frontside electrode sites was measured again (purple curve in Fig. 7d), which showed recovery of muscle fatigue but was still much lower than the measurement in the beginning. Thus, frontside electrode sites can stimulate frontside motoneurons, but can barely stimulate backside motoneurons. It proves our prediction of independent activation of motoneurons using two-side polyimide electrode sites.

**A theoretical explanation of the difference in excitability of motoneurons and muscle fibers.** There are two types of excitable cells in the muscle tissue: motoneurons and muscle fibers. They show distinctive excitability properties in electrical stimulation. In our model, these differences can be theoretically explained and proved by in vivo experiments. The experiments targeting motoneurons and muscle fibers are conducted on healthy muscle and denervated muscle, respectively.

Here, two distributed circuits are built to model the healthy muscle and denervated muscle. Since the myelin sheath is proposed as an inductor, the equivalent circuit of a myelinated motoneuron is modeled as a RLC component (Fig. 8a), while an unmyelinated muscle fiber is modeled as a RC component (Fig. 8b). To make a fair comparison, the block representing the target motoneuron or muscle fiber is placed at the same position in the circuit network. The modeling parameters are set based on two reasonable considerations as follows:

1. Due to the lack of myelin sheath, muscle fibers have a larger exposed cell membrane surface and more leakage channels, which can be modeled as a larger capacitor ($C_2 \gg C_1$) and a lower leakage resistor ($RP_2 \ll RP_1$), respectively.

2. It has been reported that the conductance and gating properties of the sodium channels in motoneuron and muscle fiber are nearly the same[22], and muscle fiber sodium channels only require a slightly more negative potential to activate than motoneuron sodium channels[23][24]. Therefore, in our modeling, the motoneurons and muscle fibers share the same parameters for probability calculus, among which the most important is the same threshold voltage, $V_{Threshold}$.

Firstly, the voltage waveforms are compared when the same negative monophasic current pulse (100 μs pulse width) is applied (Fig. 8c1). The voltage waveform of the RC component shows a typical charging and discharging curve as a capacitor, while the voltage waveform of the RLC component shows a voltage oscillation, resulting in a much higher peak amplitude. Thus, to exceed the threshold voltage and activate the muscle, the threshold current required for RLC component is much lower than RC component. The probability mapping curves by changing current amplitude in Fig. 8c2 show the difference in threshold current. The probability mapping curves with increasing pulse width in Fig. 8c3 show that RC circuit requires a much higher current to achieve the same force. This phenomenon is also observed in previous report[25] and confirmed by our in vivo experiments (Fig. 8c4). To achieve the similar force output, healthy muscle requires only 1 mA while the denervated muscle requires 10 mA.

Then the changing trend of the voltage waveforms with increasing pulse width is compared in Fig. 8d1 and Fig. 8d2. The voltage waveforms of the RC component show a simple charging and discharging process as a capacitor. Although the slope of the curve, which represents the charging rate, will vary with the circuit parameters, the general shape will not change. Thus, the effective voltage area, which is the part exceeding $V_{Threshold}$, will always have a monotonically increasing trend. As a result, the corresponding probability mapping curves (Fig. 8d3) and measured force mapping curves (Fig. 8d4) also increase with respect to pulse width monotonically. However, the changing trend of the effective voltage area of the RLC component is quite complex. Many factors, such as current amplitude, circuit parameter and extracellular environment, can induce nonlinear effect upon the effective voltage area and finally result in various probability mapping patterns, which is investigated in details in our previous C-P theory[19]. Fig. 8d2 just shows a typical voltage waveform of RLC component. With different current amplitudes, the effective voltage waveform will be completely different.

In summary, due to the lack of myelin sheath, muscle fiber is not very easily stimulated being pure RC, but because of the branching motoneurons coordinating the impulse it is able to be activated with lower current in healthy muscles. Meanwhile, the force mapping curves by changing pulse width in muscle fiber stimulation always follow a monotonically increasing trend while the force mapping curves of motoneuron stimulation will have a complex pattern[19].

**Discussion**

**A static system-level distributed-parameter circuit to study electrical stimulation on muscle tissue.** In this paper, instead of treating the muscle tissue as a black box and trying to get some simplified descriptions about the relationships between input and output, we thoroughly studied the interaction between the stimulation electrodes and muscle tissue. This interaction is highly affected by all parameters in the system, including the motoneuron-electrode position, input current waveforms (amplitude, pulse width, frequency, and polarity), circuit properties of the tissue components (motoneuron, muscle fiber, and extracellular medium), and the ion channel properties ($\alpha$, $\beta$, and $V_{Threshold}$). Therefore, a simplified general description of the relationship between input and output does not exist. No matter what kind of simplified conclusions are summarized from the experiment observations, exceptions will always exist.

Here, we demonstrate a modified distributed-parameter circuit to study a specific electrode implantation: intramuscular implantation of a double-side multiple-channel polyimide electrode. There are many electrode designs for muscle stimulation, including epimysial electrode and skin surface electrode. These electrodes also come in different structures, like flexible strip electrodes, wire electrodes and large-area patch electrodes. By including circuit components representing additional tissue layers (fat layer, skin) and properly assigning the motoneuron-electrode positions, modified distributed-parameter circuit can be applied to study all these implantations.

**A potential solution to achieve sustained force output with reduced muscle fatigue in electrical stimulation.** Our distributed-parameter circuit can provide guidance to solve practical issues in electrical stimulation. One example is to use double-side electrode stimulation for less muscle fatigue. The goal of Functional Electrical Stimulation (FES) is to achieve meaningful functions using precisely controlled electrical stimulation. However, electrical stimulation causes muscle fatigue much more easily than the voluntary controlled movements. The fast kick-in of muscle fatigue is an obstacle to achieve some meaningful functions, such as holding and gripping. In this paper, we have proved the independent recruitment of motoneurons using double-side polyimide electrode. Even when the frontside motoneurons are severely fatigued with the frontside electrodes delivering electrical stimulation, the backside motoneurons are barely affected and can still generate large force with the backside electrodes. The independent recruitment of motoneurons results in independent fatigability within a muscle. If we expand this concept to a meaningful function, such as a holding movement, the frontside and backside electrodes can be controlled to alternatively deliver the electrical stimulation (Fig. 9). By alternating the side of electrode sites during a long stimulation, the muscle fibers and motoneurons on one side can rest and recover when the other side is recruited for force output.

**Potential applications of distributed-parameter circuit model to study electrical stimulation on other neural tissue.** Apart from muscles, this distributed-parameter circuit can also be applied to other excitable tissues, including the Peripheral Nervous System (PNS) and the Central Nervous System (CNS). In the distributed-parameter circuit description for the PNS and CNS, the muscle fibers will be removed, and much more neuron distribution will be added. Further study using our distributed-parameter circuit will help to elucidate important issues in the PNS and CNS electrical stimulation. An example is to use high frequency localized electrical stimulation for selective activation. Selective stimulation of different nerve fascicles on the sciatic nerve can enable functional activation of various muscle groups. In addition, localized stimulation in the spinal cord can eliminate unwanted disturbance on other functional areas.

In the previous research, without this system-level model, researchers tended to use experiment observations to judge the efficiency of electrical stimulation. Now we know the stimulation effect under a specific condition is affected by so many parameters in the system. Since in the way of experiment observations, only a certain range of parameters can be applied, this may give rise to a segmented conclusion about this specific stimulation. This segmented conclusion may neglect the complicated interaction between the stimulation electrodes and excitable tissue. Here, we take the study of tripolar stimulation on the sciatic nerve as an example, to demonstrate how segmented conclusions can arise when only parameters of a certain range are applied during the experiment observations. In Fig. 10, the two outer electrode sites are connected to the positive terminal of the current source, and the middle electrode site is connected to the negative terminal. Positive monophasic input current is delivered using this tripolar structure. With the knowledge about voltage waveform polarity in Fig. 6, we now know that it is the middle electrode site which is delivering negative input current that stimulates the nerve axons to generate action potentials. When these action potentials travel to the electrode site located at the lower stream of the nerve axons, they will be distorted by the external stimulation at that location. Whether these action potentials can successfully pass through the lower-stream electrode site depends on the external stimulation at that location: they can pass through when the amplitude of the external stimulation is smaller than the action potential (Fig. 10a), and they will be blocked when the amplitude of the external stimulation is larger (Fig. 10b). However, it is also possible that when the action potentials arrive at the lower-stream electrode site, the external stimulation of a short pulse width has already finished. Then, in this case, the action potentials will pass through at both small and large amplitude (Fig. 10c, Fig. 10d). Thus, if we use force measurement as a quantification of the action potentials passing through the lower-stream electrode site, the force curve with respect to the current amplitude will show totally different patterns when different pulse width is used.

**A new explanation of cathodal make, cathodal break, anodal make, and anodal break stimulation phenomena.** It has been observed that cardiac tissue can be electrically stimulated with the onset (make) or termination (break) of an input current that is delivered with either a negative (cathodal) or positive (anodal) electrode[26][27]. These phenomena can be easily explained with our model. In the case of unipolar stimulation configuration, where only one electrode delivers input current, the effectiveness of cathodal make can be intuitively understood, as it will depolarize the tissue surrounding the electrode. However, considering the dramatic resonance caused by the RLC component, cathodal break, anodal make, and anodal break may also generate voltage waveforms with a certain area sufficiently negative to exceed the voltage threshold.

**The gap between static and dynamic distributed-parameter circuit model.** Inherited and expanded from the C-P theory, we still assigned static values to parameters of both the circuit components (resistance, inductance, and capacitance) and probability calculation ($\alpha$, $\beta$ and $V_{Threshold}$). By using static parameters of circuit components, we indeed assumed that the electrical properties of the muscle tissue remain unchanged during applied electrical stimulation. Similarly, by using static parameters of probability calculation, we assumed that external electrical stimulation won't affect the resting potential and threshold voltage of motoneurons. This adoption of static parameters limits our capability to study some specific issues. One example is the quantitative description of muscle fatigue during electrical stimulation, which is characterized as change in force output. During muscle fatigue, the same current input is applied, but the output force is different. It means something must have changed, either the generated voltage waveform affected by circuit parameters, or the ion channel response to the voltage waveform which is affected by probability calculation parameters.

Another limitation caused by the adoption of static parameters is that we cannot quantitatively study how an external electrical stimulation interacts with a generated action potential. With the static circuit parameters, we can only study how the external stimulation initiates an action potential. Unlike the natural action potentials, the action potential initiated by the external electrical stimulation may be distorted by this external stimulation. For the external stimulation that lingers on after the action potential is already initiated, it will interact with the movements of local ions across the neuron membrane and result in a distorted action potential.

Thus, to quantitatively study these dynamic processes in electrical stimulation, further studies are required to establish a bridge between the parameters in our model and the dynamic change of muscle tissue.

**Reason for using voltage waveforms simulated with a distributed-parameter circuit, instead of electric field distribution simulated with the existing Finite Element Modeling (FEM).** The cross-membrane electric field opens the ion channels. Considering the neuron membrane is partially permeable to ions, charged ions will accumulate on both sides of the neuron membrane. Thus, the electric field generated by these local ions is perpendicular to the membrane. To model the effect of this perpendicular cross-membrane electric field ($E$) on the neuron, we calculate the voltage across the

capacitor ($V$) in our model. This $V$ is the path integral of the electric field $E$ in the direction perpendicular to the capacitor plates (neuron membrane) as described by the equation $\Delta V = E * \Delta d$ ($\Delta V$ is a change of $V$, $\Delta d$ is a change of membrane thickness). Thus, in our model, the voltage across the capacitor can be used to characterize the electric field perpendicular to the membrane, which really opens an ion channel. However, the electric field simulated with the existing FEM is only determined by the physical boundary condition and electrode positions, which is irrelevant to the electric field perpendicular to the neuron membrane.

In our model, by calculating the voltage across the capacitor (neuron membrane), we can model the nonlinear relationship between $V(t)$ (cross-membrane voltage at time point t) and $I(t)$ (input current at time point t). In response to an external input current $I(t)$, the local ions on both sides of the neuron membrane will move. This local ion movement is not only affected by the charging and discharging process of the neuron membrane, but also affected by the surrounding tissue environment. That's why we build a distributed-parameter circuit network to account for the influence of the other parts of the tissue. However, in the existing FEM, the tissue is modeled in a medium. In this way, the existing FEM is assuming a linear relationship between $V(t)$ and $I(t)$.

**Methods**

**Animal statement.** All experiments were conducted according to protocols approved by the Institutional Animal Care and Use Committee at the National University of Singapore.

**Preparation of denervated muscles.** Sprague-Dawley rats (around 450 g) were used for the sciatic nerve transaction. Anesthesia was induced with isoflurane. Buprenorphine was injected for pain relief before the surgery. After the rat was anesthetized, a shaver was used to gently remove the fur on the left leg. Then, the skin was disinfected with 70% ethanol wipes, and an incision was made to expose the bicep femoris muscle. An incision was then made on this bicep femoris muscle to expose the sciatic nerve. The sciatic nerve was transacted on the site before it branched into three smaller nerves. Then the bicep femoris muscle and the skin were both sutured back. Right after the surgery, enrofloxacin was injected. During the first five days after the nerve transaction surgery, buprenorphine was injected twice a day and enrofloxacin was injected once a day.

**Interface implantation.** The homemade flexible interface was folded to form a loop on the tip. A suture was threaded through this loop. Then, this suture was threaded through the center of the exposed TA muscle belly. Pulling by the suture, the interface was also threaded into the muscle belly. The interface was sutured to the muscle surface for fixation. Then, the skin was sutured back.

**Electrical stimulation.** A-M SYSTEMS model 4100 isolated high-power stimulator was used for electrical stimulation. Every one second, a train of 10 negative-positive biphasic pulses of 60 Hz was applied. In the experiment of comparing four waveforms, positive-negative biphasic pulses, negative-positive biphasic pulses, positive monophasic pulses, and negative monophasic pulses were applied.

**Force data collection and analysis.** The anesthetized rat was fixed on a stand, and the ankle of the left leg was connected to a dual-range force sensor (Vernier). This force sensor was connected to a laptop through a data acquisition device (National Instruments). LabView (National Instruments) was used for on-site result visualization during the measurements. After the measurements, MATLAB (MathWorks) was used for data analysis. The recorded forces exceeding a certain amplitude threshold were used for assessing the strength of the excitability. If the forces were below this amplitude threshold, then the muscle was considered as not activated.

**Distributed-parameter circuit simulation.** The simulation was performed on MATLAB (MathWorks). Firstly, a circuit description was performed in Simulink (MathWorks). Then, current inputs of different waveforms were recursively fed to the circuit model, and the voltage responses of the targeted RC or RLC component were collected. Lastly, these voltage responses were fed into the probability equation to calculate the probability of excitation under these current inputs. The circuit parameters and probability calculation parameters are shown in Table. 1 and Table. 2.

**Acknowledgements**

This work was supported by the following grant from the National Research Foundation: Competitive Research Project 'Peripheral Nerve Prostheses: A Paradigm Shift in Restoring Dexterous Limb Function' (NRF-CRP10-2012-01), and the grant from the HIFES Seed Funding: 'Hybrid Integration of Flexible Power Source and Pressure Sensors' (R-263-501-012-133).

We would like to thank for the experiment setup support from Han Wu, Li Jing Ong and Wendy Yen Xian Peh. We also would like to thank for the animal experiment support from Gammad Gil Gerald Lasam.

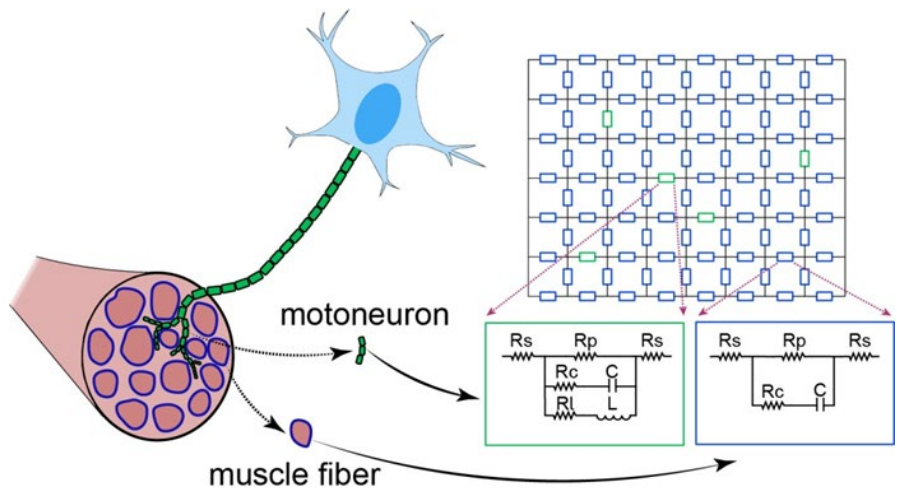

**Figure 1| Distributed-parameter circuit simulation concept.** Rs represents resistor of extracellular media. RLC components (green blocks) represent motoneurons distributed in the healthy innervated muscle tissue. RC components (blue blocks) represent muscle fibers.

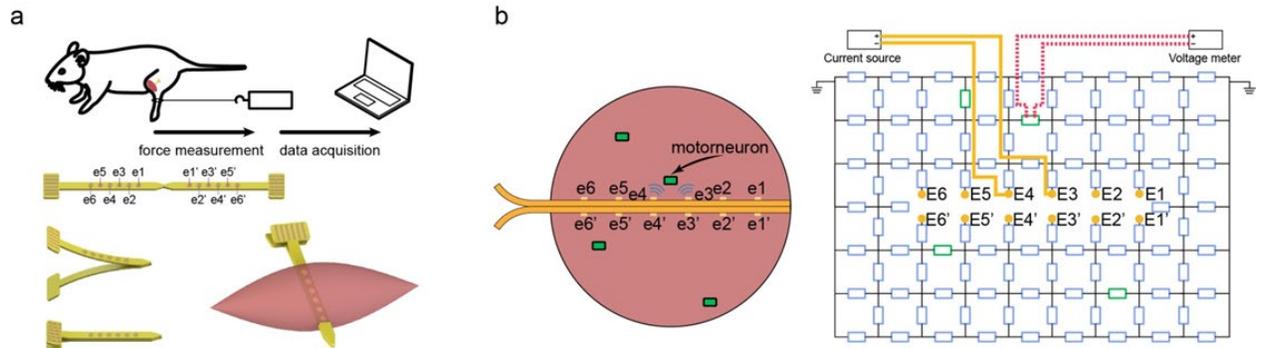

**Figure 2| In vivo measurement setup and the corresponding distributed-parameter circuit concept.** (**a**) Our home-made double-side multiple-channel polyimide electrode is sutured in the muscle belly, transversal to muscle fibers of the Tibialis Anterior (TA) muscle. When electrical stimulation is delivered to the TA muscle, the leg can freely kick forward. Force is measured from the ankle. (**b**) The corresponding distributed-parameter circuit with broken connections between two sides of electrode sites representing the insulating polyimide substrate, and RLC components (green blocks) representing distributed motoneurons.

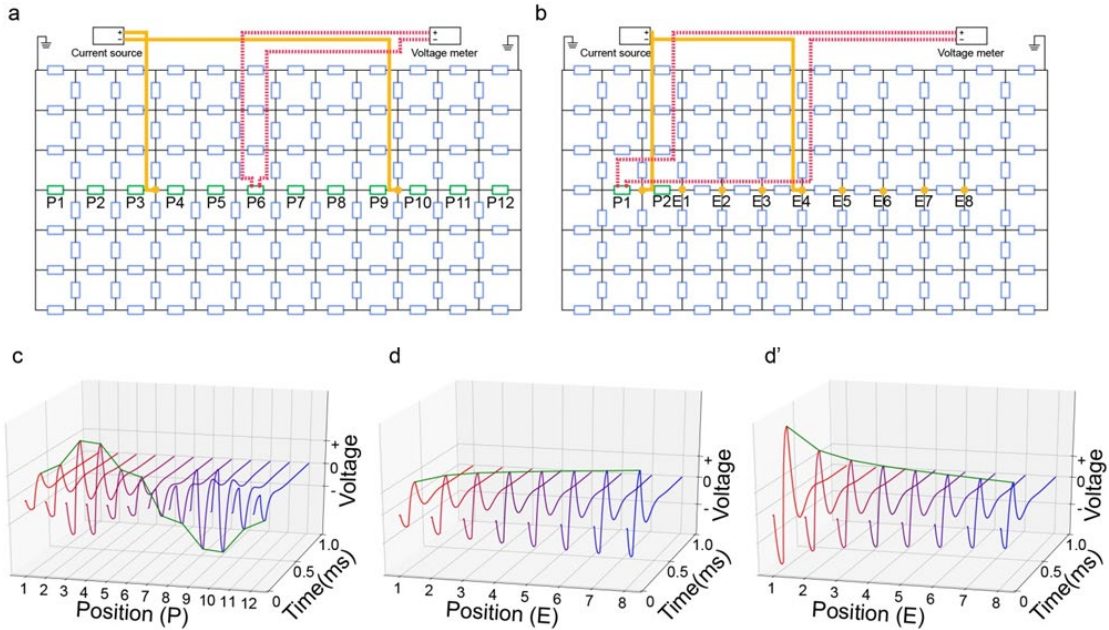

**Figure 3| Influence of motoneuron-electrode position on voltage waveform.** (**a**) Distributed-parameter circuit of fixed electrode site location and changing targeted motoneuron positions (green blocks P1-P12). (**b**) Distributed-parameter circuit of fixed positive electrode site location and changing negative electrode site locations (E1-E8). Two targeted motoneuron positions (P1, P2) are studied, P2 is between the electrode sites, and P1 is outside. (**c**) Voltage waveforms on motoneuron positions in (**a**). (**d, d'**) Voltage waveforms on P1 (**d**) and P2 (**d'**) with negative electrode positions in (**b**).

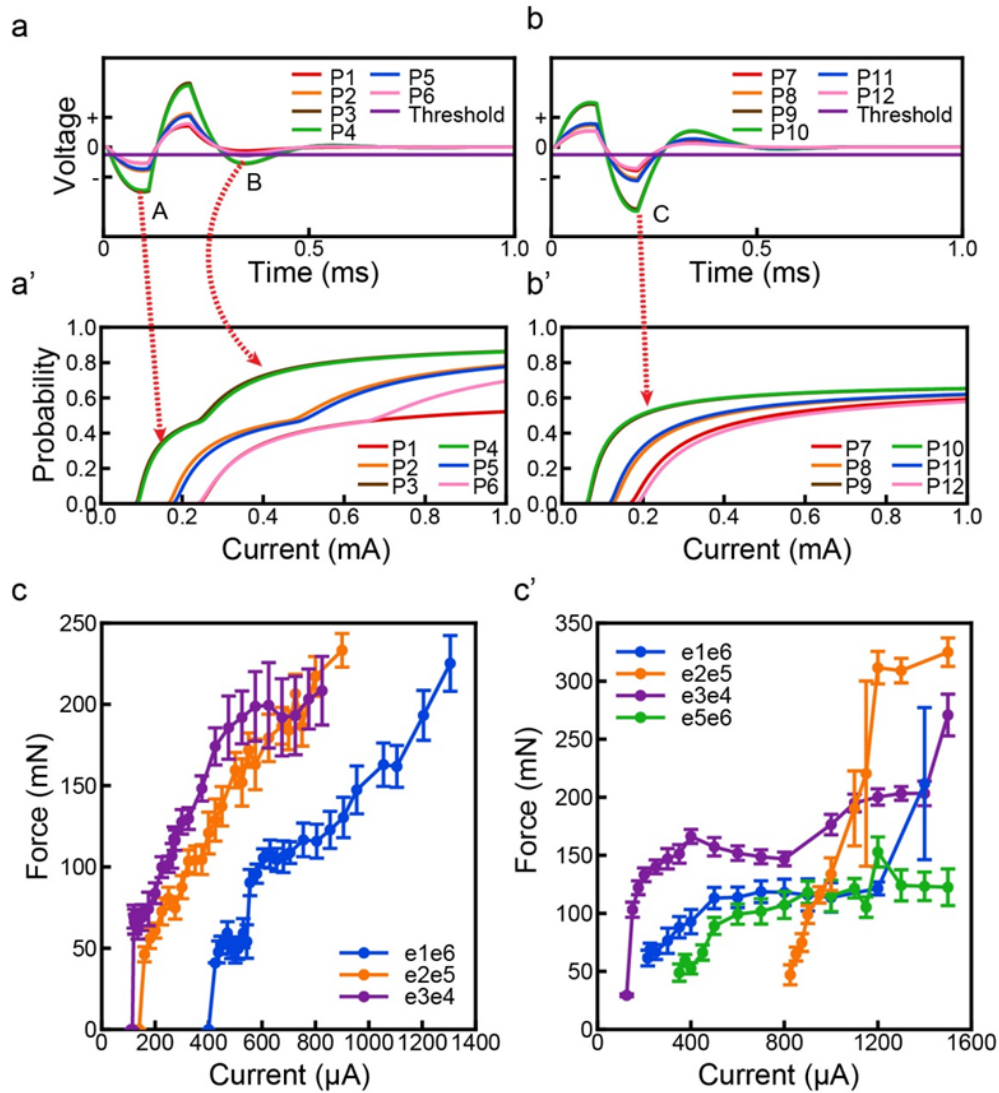

**Figure 4| Force mapping curves with multiple curvatures can be explained with the voltage waveforms.** (**a, a'**) Voltage waveforms (**a**) and corresponding probability curves (**a'**) of P1-P6 in Fig. 3c. (**b, b'**) Voltage waveforms (**b**) and corresponding probability curves (**b'**) of P7-P12 in Fig. 3c. For voltage waveforms with two areas exceeding the threshold (area A and area B), the probability curves have two corresponding curvatures. For voltage waveforms with only one area exceeding the threshold (area C), the probability curves also only show one curvature. (**c, c'**) Quantification of force at different current measured in two in-vivo experiments. Data are means ± s.d. (n=15 per group).

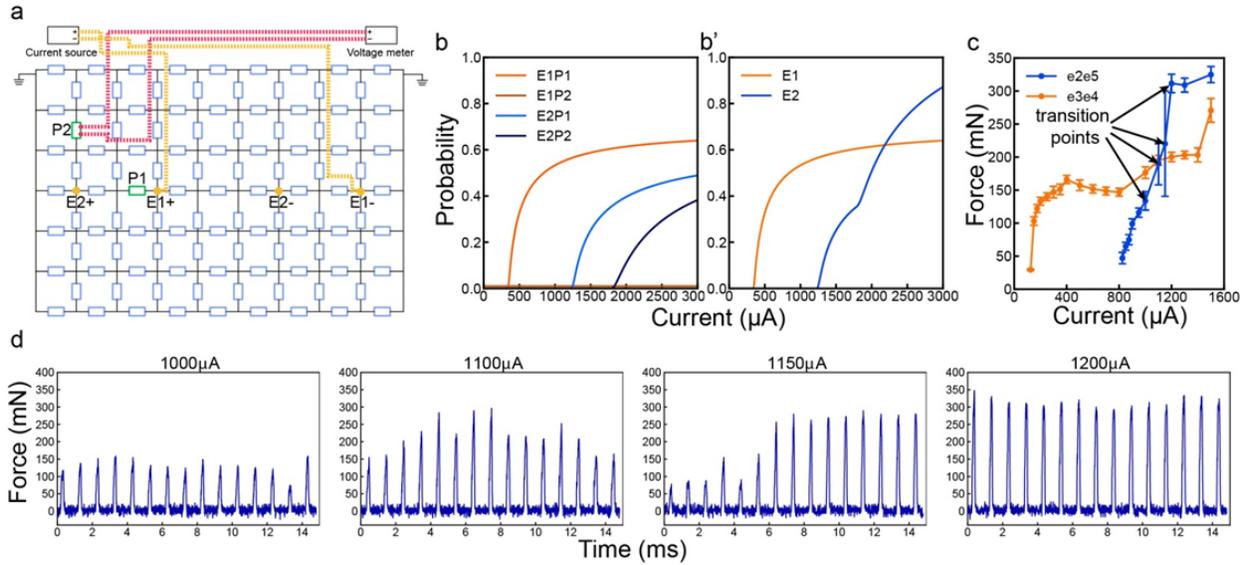

**Figure 5 | Activation of multiple groups of target motoneurons.** (**a**) Distributed-parameter circuit with two groups of electrode sites (E1+ and E1-, E2+ and E2-) and two targeted motoneuron positions (P1, P2). (**b, b'**) Simulation results. Probability curves of two groups of electrode sites on P1 and P2 separately (**b**). Probability curves of the summed stimulation effects on P1 and P2 using two groups of electrode sites (**b'**). (**c**) Quantification of force at different current. The two curves using different groups of electrodes have a crossing point. (**d**) Force profile at four current corresponding to the transition points in (**c**). Data are means ± s.d. (n=15 per group).

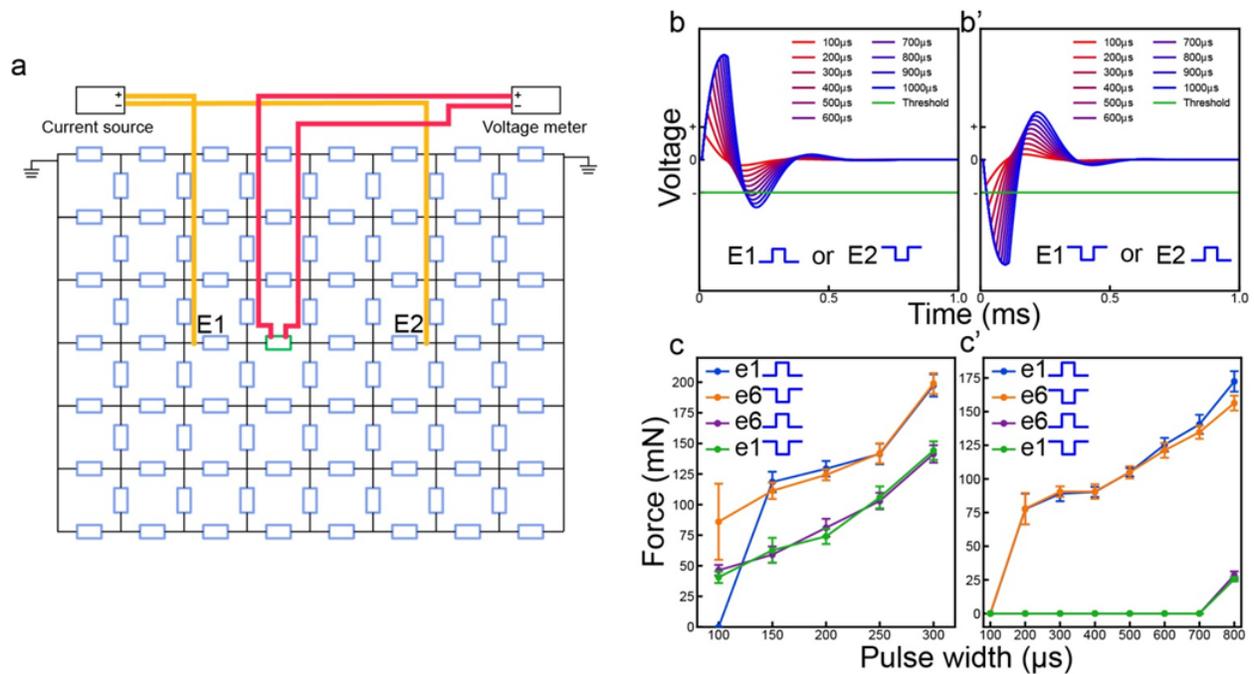

**Figure 6 | Influence of input current waveform polarity on motoneuron activation.** (**a**) Distributed-parameter circuit with two electrode sites (E1 and E2) and a targeted motoneuron (green block). (**b, b'**) Simulated voltage waveforms of different pulse width. E1 delivers positive monophasic current or E2 delivers negative monophasic current (**b**). E1 delivers negative monophasic current or E2 delivers positive monophasic current (**b'**). (**c**) Quantification of force at different pulse width in two experiments. Data are means ± s.d. (n=15 per group).

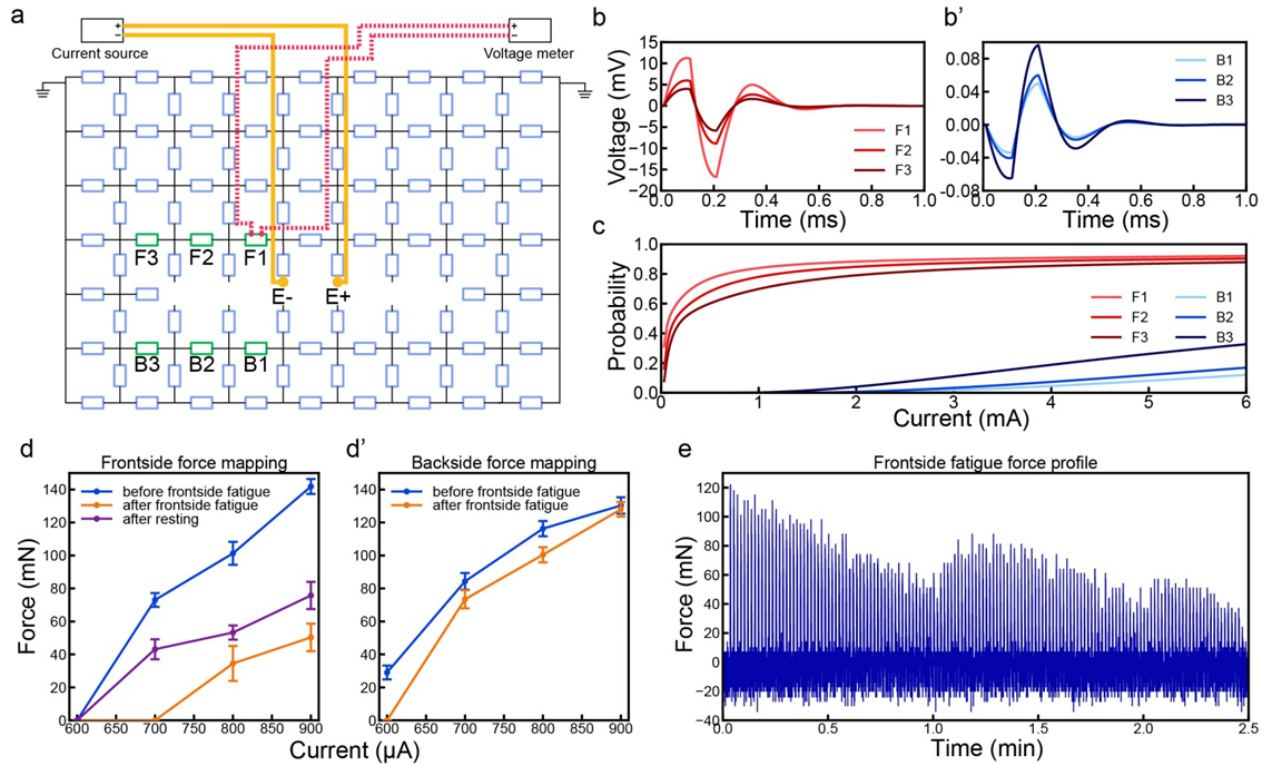

**Figure 7 | Independent activation of motoneurons using double-side polyimide electrode.** (**a**) Distributed-parameter circuit with broken connections to represent the non-conducting polyimide substrate. E- and E+ are electrode sites on the frontside of the polyimide electrode. F1, F2, F3 are motoneurons on the frontside, and B1, B2, B3 are motoneurons on the backside of the polyimide electrode. (**b, b'**) Voltage waveforms on F1-F3 (**b**) and B1-B3 (**b'**) when current input is delivered from frontside electrode sites. The scale of the front motoneuron waveforms are much larger than the back motoneuron waveforms. (**c**) The corresponding probability curves. (**d, d'**) In-vivo measurement results. Quantification of force at different current during frontside stimulation (**d**) and backside stimulation (**d'**). (**e**) Force profile of 2.5 min electrical stimulation to induce fatigue on frontside. Data are means ± s.d. (n=15 per group).

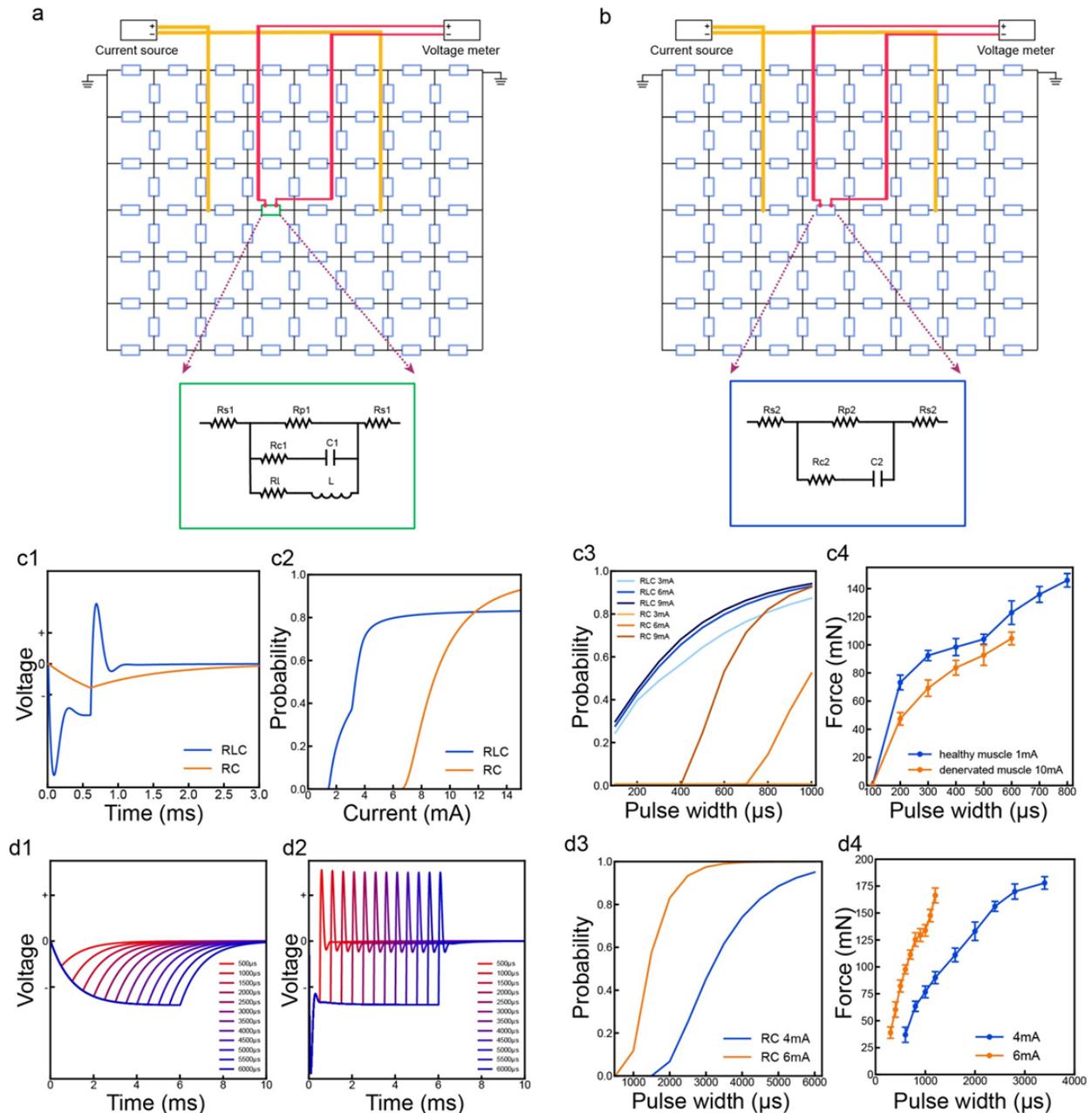

**Figure 8 | The inductive myelin sheath on motoneurons accounts for the huge difference in excitability of motoneurons and muscle fibers.** (**a**) Distributed-parameter circuit with RLC component (green block) representing the target motoneuron in healthy muscle stimulation. (**b**) Distributed-parameter circuit with RC component (blue block) representing the target muscle fiber in denervated muscle stimulation. (**c1, c2**) Simulated voltage waveforms (**c1**) and probability curves (**c2**) on RLC component and RC component (**c3, c4**) Simulated probability curves (**c3**) and quantification of force (**c4**) at small pulse width. (**d1, d2**) Simulated voltage waveforms of RC (**d1**) and RLC (**d2**) component at different pulse width (**d3, d4**) Simulated probability curve (**d3**) and quantification of force of a denervated muscle (**d4**) at large pulse width. Data are means ± s.d. (n=15 per group).

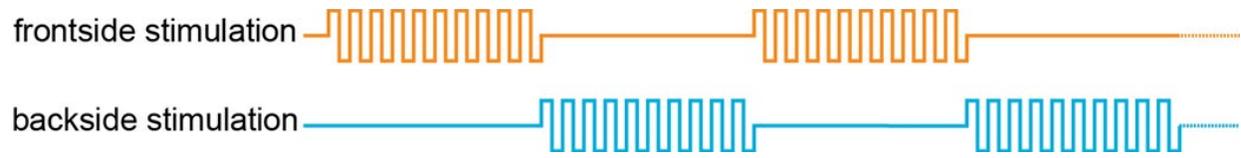

**Figure 9 | Illustration of a long-duration current waveform using double-side electrode stimulation for reduced muscle fatigue.** The frontside and backside electrode sites are controlled to alternatively deliver the stimulation current.

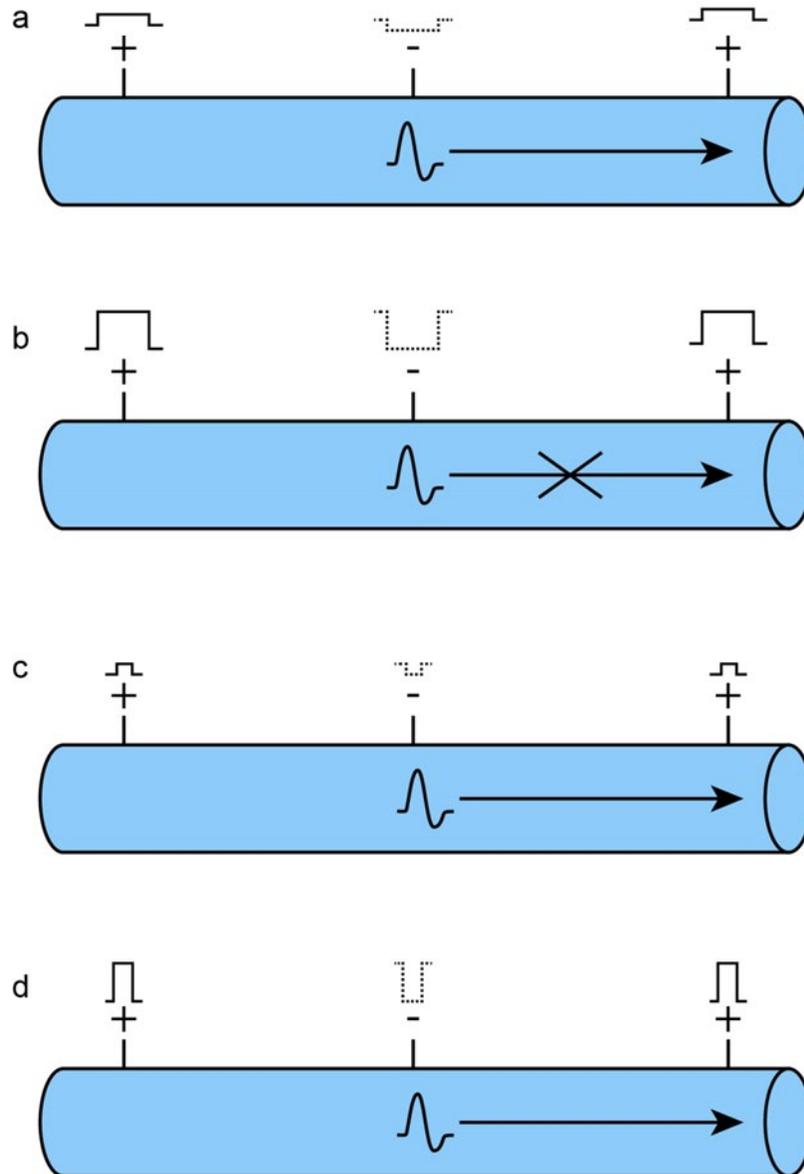

**Figure 10 | Illustration of tripolar stimulation on the sciatic nerve.** The two outer electrode sites are connected to the positive terminal of the current source, and the middle electrode site is connected to the negative terminal. A positive monophasic input current is applied using this tripolar structure. (**a, b**) Large pulse width. Action potentials generated by the negative electrode site can pass through the lower-stream positive electrode when the current amplitude is small (**a**). Action potentials cannot pass through when the current amplitude is large (**b**). (**c, d**) Small pulse width. Action potentials can pass through when the current amplitude is small (**c**) and large (**d**).

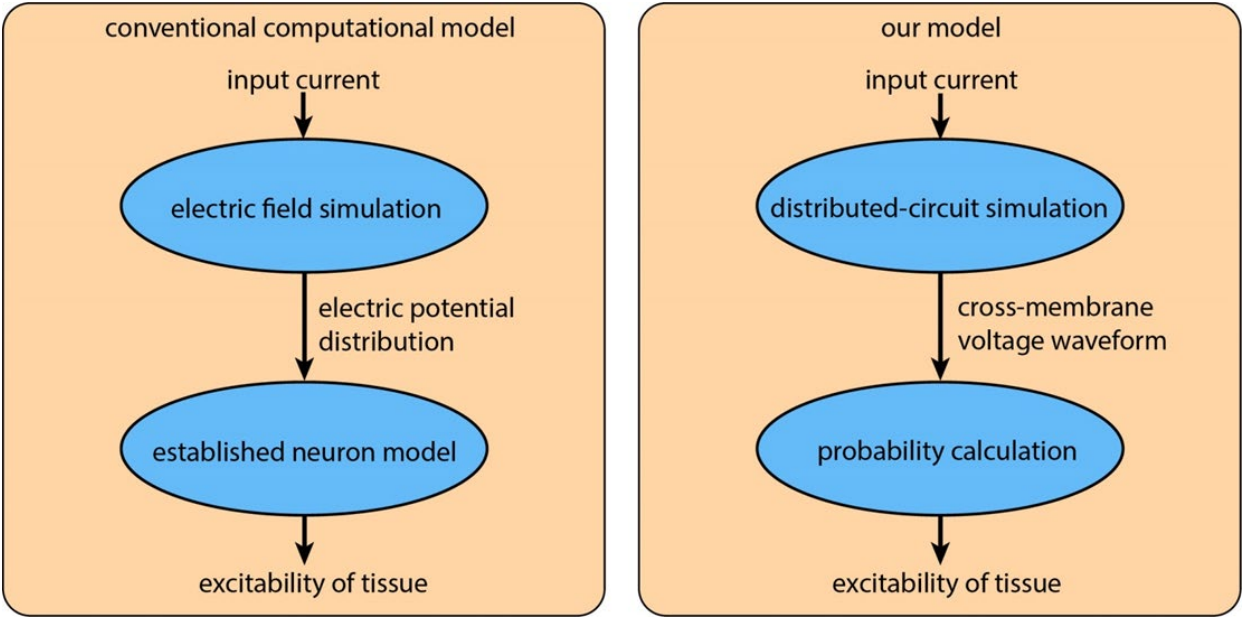

**Figure 11 | Concept of conventional computational model and our model to study excitability of tissue.**

| Rs (Ω) | Rp1 (Ω) | Rc (Ω) | C in RC (nF) | Rp2 (Ω) | RL (Ω) | C in RLC (nF) | L (H) |
|---|---|---|---|---|---|---|---|
| 100 | 100 | 100 | 1250 | 10000 | 100 | 150 | 0.02 |

Table 1 | Parameter table of components used in the distributed-parameter circuit.

| | Current range (mA) | Pulse width per phase (μs) | Waveform polarity | Threshold | α | β |
|---|---|---|---|---|---|---|
| Figure 3, Figure 4 | 0.05-1 | 100 | Negative-positive biphasic | -0.005 | 8000 | 0.001 |
| Figure 5 | 0.05-1 | 100-1000 | Negative or positive monophasic | -0.005 | 20000 | 0.001 |
| Figure 6 | 0.01-3 | 100 | Negative-positive biphasic | -0.03 | 8000 | 0.002 |
| Figure 7 | 0.03-6 | 100 | Negative-positive | -0.00002 | 8000 | 0.001 |
| Figure 8 | 0.025-15 | 500-6000 | Negative monophasic | -0.04 | 3000 | 0.005 |
| Supplementary Figure 2 | 0.05-1 | 100-1000 | Negative-positive or positive-negative biphasic | -0.005 | 20000 | 0.001 |

Table 2 | Parameter table used in distributed-parameter circuit simulation and probability calculation.

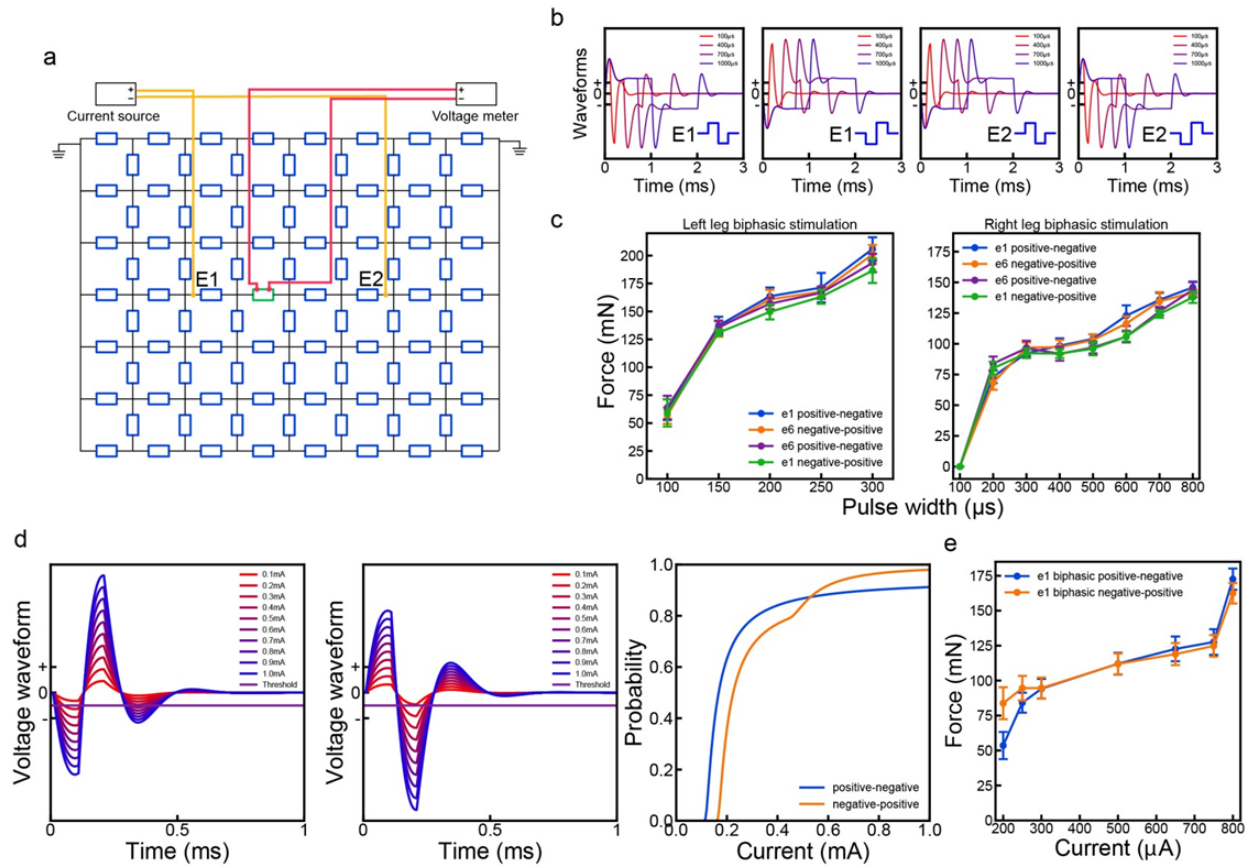

**Extended Figure 1 | Influence of input biphasic current polarity on motoneuron activation.** (**a**) Distributed-parameter circuit with two electrode sites (E1 and E2) and a targeted motoneuron (green block). (**b**) Simulated voltage waveforms of different pulse width. E1 delivering positive-negative biphasic current and E2 delivering negative-positive biphasic current share the same voltage waveform. E1 delivering negative-positive biphasic current and E2 delivering positive-negative biphasic current share the same voltage waveform. (**c**) Quantification of force at different pulse width in two experiments. According to the simulation in (**b**), the four measured recruitment curves should come in two groups (e1 positive-negative and e6 negative-positive in one group, e1 negative-positive and e6 positive-negative in another group). However, the two measurements show a single group. (**d**) To investigate the reason for showing a single group in recruitment curves, instead of two, simulation was carried out to see how the two groups of possibility curves change with stimulation current. The probability curve shows that the voltage waveform in response to positive-negative, and negative-positive current input have a slightly different changing trend with increasing current, and there is a crossing point. (**e**) Recruitment curves of e1 delivering positive-negative and negative-positive current. The two recruitment curves overlap in a large range of current. This explains why only a single group of recruitment curves is observed in (**c**).

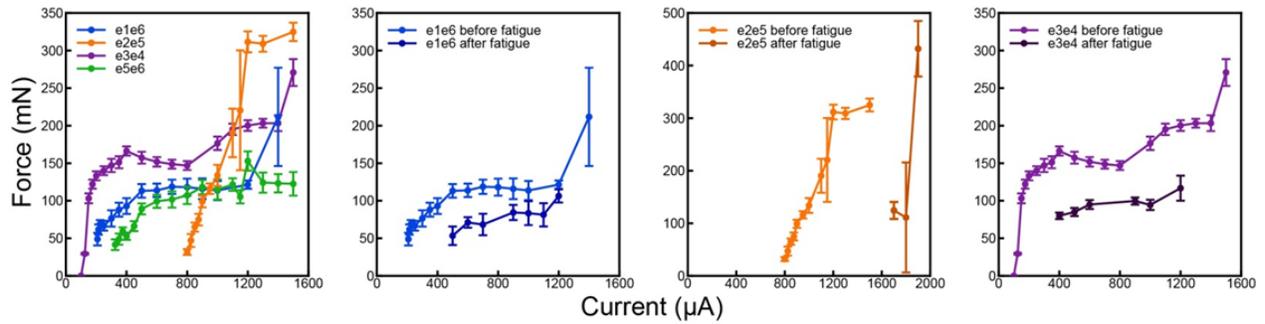

**Extended Figure 2 | Shift of force mapping curves due to muscle fatigue.** Quantification of force at different current in the first round of recruitment curve measurement (first panel). Then, the recruitment force using each electrode site combination was measured again. For each electrode site combination, the recruitment curve shifted towards larger current and smaller force output, due to accumulated muscle fatigue. Data are means ± s.d. (n=15 per group).

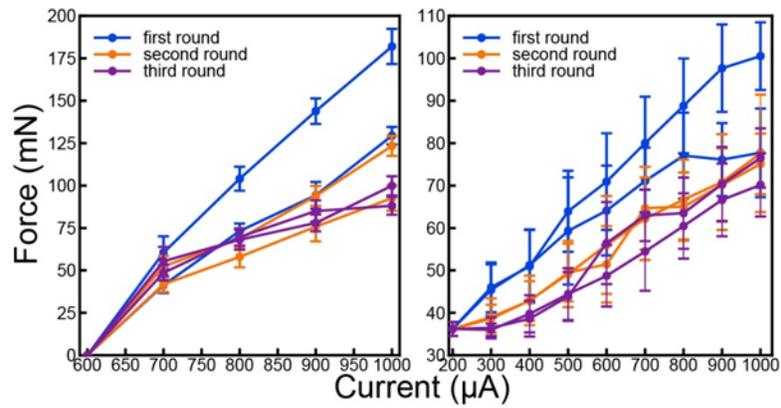

**Extended Figure 3 | Shift of recruitment curves due to muscle fatigue.** In both testing, three rounds of recruitment measurement were carried out. In each round, the current was changed from large to small, and then small back to large. In both testing, the recruitment curves shift towards smaller force output, due to accumulated muscle fatigue.

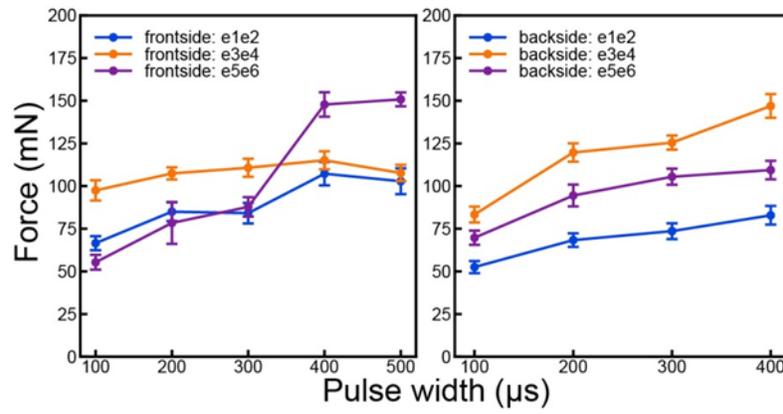

**Extended Figure 4 | Recruitment curves of denervated muscles using 10 mA current.** The recruitment curves were measured on denervated muscle, using both sides of the polyimide electrode.

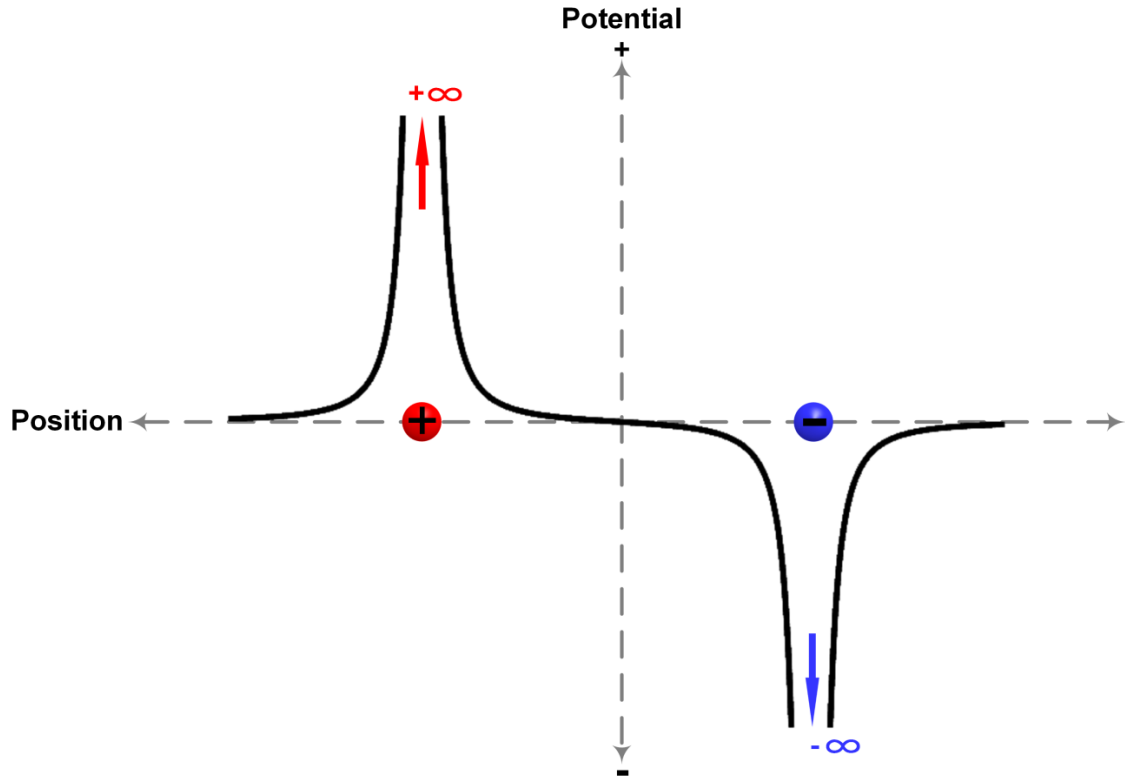

**Extended Figure 5 | Electric potential distribution generated by two point charges.** Electric potential is calculated for positions in the line of the two charges. The amplitude of the generated electric potential is the largest near the two charges. The polarity of the electric potential changes at the center of the two charges. This distribution can help to understand the change of voltage waveforms generated at different location in our model.